\newcommand{\AP}[3]{Ann.\ Phys.\ {\bf #1},\ #2 (#3)}
\newcommand{\NPA}[3]{Nucl.\ Phys.\ {\bf A#1},\ #2 (#3)}
\newcommand{\NPB}[3]{Nucl.\ Phys.\ {\bf B#1},\ #2 (#3)}

\newcommand{\PLB}[3]{Phys.\ Lett.\ B\ {\bf #1},\ #2 (#3)}
\newcommand{\PR}[3]{Phys.\ Rep.\ {\bf #1},\ #2 (#3)}
\newcommand{\PRL}[3]{Phys.\ Rev.\ Lett.\ {\bf #1},\ #2 (#3)}

\newcommand{\PRC}[3]{Phys.\ Rev.\ C\ {\bf #1},\ #2 (#3)}
\newcommand{\PRD}[3]{Phys.\ Rev.\ D\ {\bf #1},\ #2 (#3)}
\newcommand{\JPG}[3]{J.\ Phys.\ G\ {\bf #1},\ #2 (#3)}

\newcommand{\ZPC}[3]{Z.\ Phys.\ C\ {\bf #1},\ #2 (#3)}

\newcommand{\EPJA}[3]{Eur.\ Phys.\ J.\ A\ {\bf #1},\ #2 (#3)}
\newcommand{\PTP}[3]{Prog.\ Theo.\ Phys.\ {\bf #1},\ #2 (#3)}

\newcommand{\diracslash}[1]{#1\llap{/\kern2pt}}

\newcommand{\be}{\begin{equation}}
\newcommand{\ee}{\end{equation}}
\newcommand{\bea}{\begin{eqnarray}}
\newcommand{\eea}{\end{eqnarray}}
\newcommand{\ba}[1]{\begin{array}{#1}}
\newcommand{\ea}{\end{array}}

\newcommand{\eqrftw}[2]{Eqs.\ (\ref{#1}) and (\ref{#2})}

\documentclass[prd,aps,floats,nofootinbib,tightenlines,showpacs]{revtex4}
\usepackage{graphicx}
\begin{document}

\title {Color superconductivity with determinant interaction
in strange quark matter}
\author{Amruta Mishra}
\affiliation{Department of Physics, Indian Institute of Technology, New 
Delhi-110016,India}
\email{amruta@physics.iitd.ac.in}
\affiliation{Frankfurt Institute for Advanced Studies,
Universit\"at Frankfurt, D-60438 Frankfurt, Germany}
\email{mishra@th.physik.uni-frankfurt.de}

\author{Hiranmaya Mishra}
\email{hm@prl.ernet.in}

\affiliation{Theory Division, Physical Research Laboratory,
Navrangpura, Ahmedabad 380 009, India}

\date{\today} 

\def\be{\begin{equation}}
\def\ee{\end{equation}}
\def\bearr{\begin{eqnarray}}
\def\eearr{\end{eqnarray}}
\def\zbf#1{{\bf {#1}}}
\def\bfm#1{\mbox{\boldmath $#1$}}
\def\hf{\frac{1}{2}}
\begin{abstract}
We investigate the effect of six fermion determinant interaction
on color superconductivity as well as on chiral symmetry breaking.
Coupled  mass gap equations and the superconducting gap equation
are derived through the minimisation of the thermodynamic potential.
The effect of nonzero quark -- antiquark condensates on the superconducting gap
is derived. This becomes particularly relevant for 
the case of 2-flavor superconducting matter with unpaired 
strange quarks in the diquark channel. While the effect of six fermion
interaction leads to an enhancement of u-d superconductivity,
due to nonvanishing strange quark--antiquark condensates, 
such an enhancement will be absent at higher densities
for u-s or d-s superconductivity due to early
(almost) vanishing of light quark antiquark condensates.

\end{abstract}


\maketitle

 \section{Introduction}

Cold and dense quark matter can be a color superconductor.
At sufficiently high baryon densities, when nucleons 
get converted to quark matter, the resulting quark matter is
expected to be  in one kind 
or the other of the many different possible color
superconducting phases at low enough temperatures \cite{review}. The rich
phase structure is essentially due to the fact that the quark--quark interaction
is not only strong and attractive in many channels, but also many degrees of
freedom become possible for quarks like color, flavor and spin.
Studying the properties of color superconducting phases in heavy ion
 collision experiments seems unlikely
in the present accelerators as one cannot avoid producing large entropy
per baryon in heavy ion collisions and hence cannot produce the dense and
cold environment that is needed to support the formation of superconducting 
phases.  However, in the future accelerator facility planned at GSI for 
compressed baryonic matter experiments, one possibly can hope for 
observing fluctuations signifying precursory phenomena of color 
superconducting phase \cite{kunihiroo}.

        On the other hand, it is natural to expect some color superconducting
phases to exist in the core of compact stars where the densities are above
nuclear matter densities and temperatures are of the order of tens of KeV.
However, to consider quark matter for neutron stars, color and
electrical charge neutrality conditions need to be imposed for the bulk
quark matter. Such an attempt has been made 
in Ref.\cite{krisch} as well as in Ref. \cite{reddy}, where the 
lighter up and down quarks form the two flavor color superconducting (2SC)
matter while the strange quarks do not participate in pairing. 
A model independent analysis was done  in Ref. \cite{krisch}
that is valid in the limit $m_s<<\mu$ and $\Delta\sim m_s^2/\mu$, where,
 $\Delta$ is the pairing gap and $\mu$ is the quark chemical potential.
It has been shown, based upon the comparison of
free energies that such a two flavor color superconducting phase
would be absent in the core of neutron stars \cite{krisch}.
Within Nambu Jona-Lasinio (NJL) model in Ref. \cite{reddy}
it has been argued that such conclusions are consistent except for a small
window in  density range where superconducting phase is 
possible. There have also been studies to include the possibility of mixed 
phases \cite{bubmix} of superconducting matter demanding neutral
matter on the average. Later, it was observed that the imposition of 
neutrality conditions lead to pairing of quarks with different Fermi momenta
giving rise to gapless modes \cite{igor,hmam}. Within a Nambu Jona Lasinio model,
the two flavor superconducting quark matter (2SC)  was shown to 
exhibit gapless modes (g2SC) arising due to the difference in the Fermi momenta 
of the pairing quarks, when charge and color neutrality conditions are
imposed. 
Superconducting quark matter with unpaired strange quarks (2SC+s) was shown to 
exhibit gapless superconductivity (g2SC+s) within a window of about
80 MeV in baryon chemical potential \cite{hmam}. 
Temperature effects on the gapless modes were also studied for the
two flavor quark matter in Ref.\cite{igorr,amhm5}.

 A variational approach  was used to study the chiral symmetry breaking
as well as color superconductivity for the quark matter \cite{hmam,amhm5,hmparikh}.
The calculations were carried out for the NJL model 
with the minimisation of the free energy density
to study which condensate will exist at what density.
Charge neutrality conditions were introduced through the introduction
of  appropriate chemical potentials.  We note here that
 the possibility of diquark condensates along with quark--antiquark 
condensates has also been considered in Ref.s
\cite{{berges},{mei},{blaschke},{kunihiro}}.

In the present work, we investigate the effect of six fermion determinant 
interaction on  color  superconductivity in Nambu JonaLasinio model.
Since it is not possible to have a model with same symmetries as QCD with
four quark operators alone, the so called t`Hooft term is added
to the usual four fermion interaction which is of the form
\be
{\cal L}_{det}\sim 
det_f[\bar\psi(1+\gamma_5)\psi]
+det_f[\bar\psi(1-\gamma_5)\psi]
\label{thooft}
\ee
where, determinant is in the flavor (u,d,s) space. This term respects chiral symmetry
and breaks $U(1)$  axial symmetry as in  QCD.

While studying the dense quark matter, the effect of such a term in  quark
antiquark channel has been widely used \cite{reddy,ruester,bubmix,aichlin}.
However, for color superconductivity
such a term is only considered at the four fermion level.
The effect of six fermion interaction has been considered recently in 
Ref. {\cite{steiner}}. This, however considers a different
effective six fermion interaction than as given in 
Eq.(\ref{thooft}). We shall treat both
chiral symmetry breaking as well as color superconductivity
in the same footing using the determinant interaction as in 
Eq.(\ref{thooft}). With the variational ansatz considered, we show that
such a term gives a nonzero contribution to the free energy and
at higher densities, it is of similar or even larger  magnitude than 
the contributions from the quark--antiquark condensates.

We organize the paper as follows. In the following subsection, we discuss the ansatz
state with the quark--antiquark as well as the diquark condensates 
\cite{amhm5}
.
In section \ref{evaluation},  we consider the Nambu Jona-Lasinio model
Hamiltonian with the determinant interaction
and calculate its expectation value with respect to the
ansatz state to compute the energy density as well as the 
thermodynamic potential. We minimise the thermodynamic potential
to calculate all the ansatz functions and the resulting mass as well as
superconducting gap equations here. In section \ref{results} we discuss
the results of the present investigation. Finally we summarise 
and conclude in section \ref{summary}.
\subsection{ The ansatz for the ground state}

%
We shall use the notations and conventions of Ref. \cite{amhm5,hmam} and
recapitulate
briefly the construction of the variational ansatz for the ground state.
We take  it  as a squeezed
coherent state involving quark--antiquark as well as diquark condensates
as given by
\cite{amhm5,hmam,hmparikh} 
\begin{equation} 
|\Omega\rangle= {\cal U}_d|vac\rangle={\cal U}_d{\cal U}_Q|0\rangle.
\label{u0}
\end{equation} 

Here, ${\cal U}_Q$ and ${\cal U}_d$ are unitary operators creating
quark--antiquark and diquark pairs respectively. Explicitly,
the operator, ${\cal U}_Q$ is given as
\begin{equation}
{\cal U}_Q=\exp\left (
\int q^{0i}(\zbf k)^\dagger(\bfm {\sigma }\cdot\zbf k) h_i(\zbf k)
 \tilde q^{0i} (\zbf k)d\zbf k-h.c.\right ).
\label{u1}
\end{equation}
 
\noindent In the above,  $q^0$ ($\tilde q^0$) is the two component 
particle annihilation (antiparticle creation) opertor of the four componet
quark field operator $\psi$. The superscript $`0'$ refers to the fact
 that these operators correspond to ``free" Dirac fields 
including, in general, a current quark mass. The quark--antiquark
condensate function $h_i(\zbf k)$  is a
real function of $|\zbf k|$ which describes vacuum realignment for
chiral symmetry breaking for quarks of a given flavor $i$. We shall take 
the condensate function $h(\zbf k)$ to be the same for u and d quarks and
$h_3(\zbf k)$ as the chiral condensate function for the s-quark.
Clearly, a nontrivial $h_i(\zbf k)$ shall break the chiral
symmetry. Summation over three colors and three flavors is understood in the
exponent of ${\cal U}_Q$
in Eq. (\ref{u1}). Similarly, the unitary operator ${\cal U}_d$
describing diquark condensates is given as

\begin{equation}
{\cal U}_d=\exp(B_d^\dagger-B_d)
\label{omg}
\end{equation}
where, $B_d^\dagger$ is the pair creation operator as given by
\begin{equation}
{B}_d ^\dagger=\int \left[q_r^{ia}(\zbf k)^\dagger
r f(\zbf k) q_{-r}^{jb}(-\zbf k)^\dagger
\epsilon_{ij}\epsilon_{3ab}
+\tilde q_r^{ia}(\zbf k)
r f_1(\zbf k) \tilde q_{-r}^{jb}(-\zbf k)
\epsilon_{ij}\epsilon_{3ab}\right]
d\zbf k.
\label{bd}
\end{equation}
\noindent 
In the above, $i,j$ are flavor indices, $a,b$ are the
color indices and $r(=\pm 1/2) $ is the spin index.
The operators $q(\zbf k)$ are related to $q^0(\zbf k)$ through
the transformation $q(\zbf k)={\cal U}_Q q^0(\zbf k){\cal U}_Q^{-1}$.
As noted earlier we 
shall have the quarks of colors red and green ($a$=1,2) and 
flavors u,d ($i$=1,2) taking part in diquark condensation.
The blue quarks (a=3) do not take part in diquark condensation.
Note that we have assumed the `condensate functions' $f(\zbf k)$ to be
independent of flavor color indices. We give a post-facto justification 
for this to be that the function depends upon the {\it{ average}} energy 
and {\it average}
chemical potential of the quarks that condense and is the same for
red up and green down or green up and red down quarks, when color isospin
is unbroken.

Finally, to include the effects of temperature and density we next write
 down the state at finite temperature and density 
$|\Omega(\beta,\mu)\rangle$  taking
a thermal Bogoliubov transformation over the state $|\Omega\rangle$ 
using thermofield dynamics (TFD) as described in Ref.s \cite{tfd,amph4}.
We then have,
\begin{equation} 
|\Omega(\beta,\mu)\rangle={\cal U}_{\beta,\mu}|\Omega\rangle={\cal U}_{\beta,\mu}
{\cal U}_d{\cal U}_Q |0\rangle.
\label{ubt}
\end{equation} 
where ${\cal U}_{\beta,\mu}$ is given as
\begin{equation}
{\cal U}_{\beta,\mu}=e^{{\cal B}^{\dagger}(\beta,\mu)-{\cal B}(\beta,\mu)},
\label{ubm}
\end{equation}
with 
\begin{equation}
{\cal B}^\dagger(\beta,\mu)=\int \Big [
q_I^\prime (\zbf k)^\dagger \theta_-(\zbf k, \beta,\mu)
\underline q_I^{\prime} (\zbf k)^\dagger +
\tilde q_I^\prime (\zbf k) \theta_+(\zbf k, \beta,\mu)
\underline { \tilde q}_I^{\prime} (\zbf k)\Big ] d\zbf k.
\label{bth}
\end{equation}
In Eq.(\ref{bth}) the ansatz functions $\theta_{\pm}(\zbf k,\beta,\mu)$
will be related to quark and antiquark distributions, and, the underlined
operators are the operators in the extended Hilbert space associated with
thermal doubling in TFD method. In Eq.(\ref{bth}) we have suppressed
the color and flavor indices on the quarks as well as the functions
$\theta(\zbf k,\beta,\mu)$.

 All the functions in the ansatz in Eq.(\ref{ubt})
are to be obtained by minimising the
thermodynamic potential.
We shall carry out this minimisation
in the next section.

\section{Evaluation of thermodynamic potential and gap equations }
\label{evaluation}
As discussed earlier we shall consider here 3-flavor 
Nambu Jona Lasinio model including the determinant interaction
given by equation (\ref{thooft}). The
Hamiltonian is then given as
\bearr
{\cal H} & = &\psi^ \dagger(-i\bfm \alpha \cdot \bfm \nabla
+\gamma^0 \hat m )\psi
-G_s\sum_{A=0}^8\left[(\bar\psi\lambda^A\psi)^2-
(\bar\psi\gamma^5\lambda^A\psi)^2\right]\nonumber\\
&-&G_D(\bar\psi\gamma^5\epsilon\epsilon^b\psi^C)
(\bar\psi^C\gamma^5\epsilon\epsilon^b\psi) 
+K\left[{ det_f[\bar\psi(1+\gamma_5)\psi]
+det_f[\bar\psi(1-\gamma_5)\psi]}\right]
\label{ham}
\eearr
where $\psi ^{i,a}$ denotes a quark field with color `$a$', 
$(\rm a=r,g,b)$ and flavor `$i$',
 $(\rm i=u,d,s)$ indices. The matrix of current quark masses is given by
$\hat m$=diag$_f(m_u,m_d,m_s)$ in the flavor space.
As noted earlier, we  shall assume isospin
symmetry with $m_u$=$m_d$.  
In Eq.(\ref{ham}), $\lambda^A$, $A=1,\cdots 8$ denote the Gellman matrices
acting in the flavor space and
$\lambda^0 = \sqrt{\frac{2}{3}}\,1\hspace{-1.5mm}1_f$,
$1\hspace{-1.5mm}1_f$ as the unit matrix in the flavor space.
The four point interaction term $\sim G_s$ is symmetric in $SU_V(3)\times
SU_A(3)\times U_V(1)\times U_A(1)$. In contrast, the determinant term
$\sim K$ which for the case of three flavors generates a six point
interaction which breaks $U_A(1)$ symmetry. If the diquark and the mass terms are
neglected, the overall symmetry is $SU_V(3)\times SU_A(3) \times U_V(1)$. This
spontaneously breaks to $SU_V(3) \times U_V(1)$ implying
the conservation of the baryon number and the flavor number. The current
quark mass term introduces additional explicit breaking of chiral symmetry
and the axial flavor current is not completely conserved.

The third term in Eq.(\ref{ham}) describes a scalar diquark interaction
in the color antitriplet and flavor antitriplet channel. We have not included,
in the present investigation,
a pseudoscalar term with the same coupling which leads to Goldstone mode
condensation.
Such a  form of Lagrangian can arise e.g. by Fiertz
transformation of a four point current current interaction having 
quantum numbers of single-gluon exchange \cite{ebert}. In that case 
the diquark coupling $G_D$ is related to the scalar coupling
as $G_D=0.75G_s$.

Using the variational ansatz state in Eq.(\ref{ubt}) 
one can calculate the expectation values
of various operators \cite{hmam}.
We evaluate the expectation values
\begin{equation}
\langle \Omega(\beta,\mu)
 |\tilde\psi_\alpha^{ia}(\zbf k)\tilde\psi ^{jb}_\beta(\zbf k')^{\dagger}
|\Omega(\beta,\mu)\rangle
=\delta^{ij}\delta^{ab}
\Lambda_{+\alpha\beta}^{ia}(\zbf k,\beta,\mu)\delta(\zbf k-\zbf k'),
\label{psipsidb}
\end{equation}
and,
\begin{equation}
\langle \Omega(\beta,\mu)
|\tilde\psi_\beta(\zbf k)^{ia\dagger}\tilde\psi_\alpha^{jb}(\zbf k')
|\Omega(\beta,\mu)\rangle
=\delta^{ij}\delta^{ab}
\Lambda_{-\alpha\beta}^{ia,jb}(\zbf k,\beta,\mu)\delta(\zbf k-\zbf k'),
\label{psidpsib}
\end{equation}
where,
\begin{equation}
\Lambda_\pm^{ia}(\zbf k,\beta,\mu)
=\hf\left[1\pm \left( F_1^{ia}(\zbf k)
-F^{ia}(\zbf k)\right)\pm \big(\gamma^0\cos\phi^i (\zbf k)\big)
+\bfm\alpha\cdot\hat\zbf k\sin \phi^i
(\zbf k)\big)\big(1-F^{ia}(\zbf k)-F_1^{ia}(\zbf k)\big)
\right].
\label{prpb}
\end{equation}
In the above, 
$\tilde\psi(\zbf k)$ is the Fourier transform of $\psi(\zbf x)$ \cite{amhm5}.
The effect of the diquark condensates and their temperature and/or density
dependences are encoded in the functions $F^{ia}(\zbf k)$ and $F_1^{ia}
(\zbf k)$ given as
\begin{equation}
F^{ia}(\zbf k)=\sin^2\theta^{ia}_-(\zbf k)+\sin^2 f
\cos\theta_-^{ia,jb}(\zbf k)
\left(1-\delta^{a3}\right),
\label{fkb}
\end{equation}
and,
\begin{equation}
F_1^{ia}(\zbf k)=\sin^2\theta^{ia}_+(\zbf k)+\sin^2 f_1
\cos 2\theta_+^{ia,jb}(\zbf k)
\left(1-\delta^{a3}\right).
\label{f1kb}
\end{equation}
We have defined $\cos 2\theta_{\pm}^{ia,jb}=1-\sin^2\theta^{ia}_{\pm}-
\sin^2\theta_{\pm}^{jb}$ with $i\neq j$ and $a\neq b$.
The $\delta^{a3}$ term indicates that the third color does not take part in 
diquark condensation. Further, we have introduced the notation for the 
quark--antiquark condensate functions as a `shift' from their vaccum values as
$\phi_i(\zbf k)=\phi_i^0(\zbf k)-2 h_i(\zbf k)$, with 
$\cot \phi^0_i(k)=m_i/\sqrt{(k^2+m_i^2)}$.

\noindent
We also have for diquark operators,

\bearr
\langle \Omega(\beta,\mu)| \psi^{ia}_\alpha(\zbf x)\psi^{jb}_\gamma
(\zbf 0)
|\Omega(\beta,\mu) \rangle
&=&-\frac{1}{(2\pi)^3}
\int e^{i\zbf k\cdot\zbf x}
{\cal {P}}_{+\gamma\alpha}^{ia,jb}(\zbf k,\beta,\mu)d\zbf k,\nonumber\\
\langle \Omega(\beta,\mu)| \psi^{ia\dagger}_\alpha(\zbf x)
\psi^{jb\dagger}_\gamma
(\zbf 0)
|\Omega(\beta,\mu) \rangle
&=&-\frac{1}{(2\pi)^3}\int e^{i\zbf k\cdot\zbf x}
{\cal {P}}_{-\alpha\gamma}^{ia,jb}(\zbf k,\beta,\mu)d\zbf k,
\label{psi}
\eearr

where, 
\bearr
{\cal{P}}_+^{ia,jb}(\zbf k,\beta,\mu)
&=&\frac{\epsilon^{ij}\epsilon^{3ab}}{4}\bigg[S^{ia,jb}(\zbf k)
\cos\left(\frac{\phi_i-\phi_j}{2}\right)\nonumber\\
&+&\left(\gamma^0
\cos \left(\frac{\phi_{i}+\phi_j}{2}\right)-\bfm\alpha\cdot
\hat\zbf k\sin\left(\frac{\phi_i+\phi_j}{2}\right)\right)A^{ia,jb}(\zbf k)
\bigg]\gamma_5 C,
\label{calpp}
\eearr
and,
\bearr
{\cal{P}}_-^{ia,jb}(\zbf k,\beta,\mu)
&=&\frac{\epsilon^{ij}\epsilon^{3ab}
C\gamma_5}{4}\bigg [S^{ia,jb}(\zbf k)
\cos\left(\frac{\phi_i-\phi_j}{2}\right)\nonumber\\
&+&
\left(\gamma^0\cos\left(\frac{\phi_i+\phi_j}{2}\right)
-\bfm\alpha\cdot \hat\zbf k\sin \left(\frac{\phi_i+\phi_j}{2}\right)
\right)A^{ia,jb}(\zbf k)
\bigg].
\label{calpm}
\eearr

\noindent Here, $C=i\gamma^2 \gamma^0$ is the charge conjugation matrix (we
use the notation of Bjorken and Drell) and the functions $S(\zbf k)$ and 
$A(\zbf k)$ are given as,
\be
S^{ij,ab}(\zbf k)=\sin\!2f(\zbf k) \cos 2\theta^{ia,jb}_-(\zbf k,\beta,\mu)
+\sin\!2f_1(\zbf k)\cos 2\theta^{ia,jb}_+(\zbf k,\beta,\mu),
\label{sk}
\ee
and,
\begin{equation}
A^{ij,ab}(\zbf k)=\sin\!2f(\zbf k)\cos 2\theta^{ia,jb}_-(\zbf k,\beta,\mu)
-\sin\!2f_1(\zbf k)\cos 2\theta_+^{ia,jb}(\zbf k,\beta,\mu),
\label{ak}
\end{equation}
These expressions are used to calculate thermal 
expectation value of the Hamiltonian and compute the thermodynamic
potential. 

Let us first concentrate on the contribution of the determinant term
to the energy expectation value. When expanded the determinant term
will have six terms, each involving three pairs of quarks of different
flavors. These are to be `contracted' among themselvs in all
possible manner, while taking expectation values.
Further, while considering quark--antiquark condensates,
it is clear that the contractions of the same color will be
dominant over that with different colors by a factor $N_c$, where
$N_c$ is the number of colors. With the present case of
two flavor superconductivity, this leads to the fact that
out of the six terms in the expansion of the determinant, only two terms
that are proportional to the strange quark--antiquark condensate
$\langle\bar s s\rangle$ will be dominant over the rest.
These latter four terms  are suppressed at least by a factor $N_c$.
The dominant  two terms are the ones involving contraction
of strange quark--antiquarks having the same color.
This simplification arises because we are considering
only u-d superconductivity here.
For the rest of our calculations, we take contributions of these two terms only.
Explicitly these two terms are given as $\sim \sum_a(\bar s\hat O^a s)\left[
(\bar u \hat O^a u)(\bar d\hat O^a d)-(\bar u \hat O^a d)(\bar d\hat O^a u)
\right]$,
where we have written ,with $a=\pm$, $\hat O^\pm=(1\pm\gamma_5)$. Using,
Eq.s (\ref{psi})--Eq. (\ref{calpm}), it is starightforward to show that, 
both the terms in the 
square bracket give identical contribution to the expectation value when
`contracted' diquarkwise, except, that the contribution of the second term 
is of opposite sign as compared to the first term due to the flavor antisymmetric
nature of the expectation values.
The determinant term
contribution is then given as

\bearr
V_{det} &=&
+K\langle{ det_f[\bar\psi(1+\gamma_5)\psi]
+det_f[\bar\psi(1-\gamma_5)\psi]}\rangle\nonumber\\
&=&-8 KI_s^{(3)}\left(I_D^2+2 I_s^{(1)}I_s^{(2)}\right).
\label{vdet}
\eearr

\noindent Here, 
\be
I_s^{(i)}=-\frac{1}{2}\langle\bar\psi^i\psi^i\rangle
=\sum_{a=1}^3\int\frac{d\zbf k}{(2\pi)^3}\cos\phi^i(1-F^{ia}-F_1^{ia})
\label{is}
\ee
 is proportional to the quark--antiquark condensate
for $i$-th flavor, and, 
\bearr
I_D&=&\frac{1}{4}\langle\bar{\psi^c}^{ia}\gamma^5\epsilon^{ij}\epsilon^{3ab}\psi^{jb}\rangle\nonumber\\
& =&\frac{1}{(2\pi)^3}\int d{\zbf k}\cos (\frac{\phi^1-\phi^2}{2})\left[
\sin 2f(\zbf k) (1-\sin^2\theta_-^1-\sin^2\theta_-^2)
+\sin 2f_1(\zbf k) (1-\sin^2\theta_+^1-
\sin^2\theta_+^2)\right]
\label{id}
\eearr
is the diquark condensate.
It is then straightforward  to calculate the expectation value of the Hamiltonian.
This can be written as
\be
\epsilon = \langle H \rangle = T +V_S +V_D +V_{det},
\ee
the various terms arising from the kinetic, the scalar,
the diquark and the determinant
interaction terms of the Hamiltonian respectively. Explicitly,
the kinetic energy part in Eq.(\ref{ham}) is given as 
\bearr
T & \equiv &
 \langle \Omega(\beta,\mu)|
\psi^\dagger
(-i\bfm \alpha \cdot \bfm \nabla +\gamma^0 m_i )\psi
| \Omega(\beta,\mu)\rangle \nonumber \\
 & = & -\frac{2}{(2\pi)^3}\sum_{i=1}^3\sum_{a=1}^3
\int d \zbf k (|\zbf k|\sin \phi^i+m_i\cos\phi_i)(1-F^{ia}-F_1^{ia}),
\label{tren}
\eearr
where, $F^{ia}$ and $F_1^{ia}$ are given by Eq.s(\ref{fkb}) and Eq.(\ref{f1kb}).

The contribution from the scalar interaction  term in Eq.(\ref{ham})
turns out to be
\begin{equation}
{V_S}\equiv -G_s \langle \Omega(\beta,\mu)|
\sum_{A=0}^8\left[(\bar\psi\lambda^A\psi)^2-
(\bar\psi\gamma^5\lambda^A\psi)^2\right]
| \Omega(\beta,\mu)\rangle
=-8 G_S\sum_{i=1,3}{I_S^i}^2,
\label{vs}
\end{equation}

where, $I_s^i$ is given in Eq.(\ref{is}).

Similarly, the contribution from the diquark interaction
from Eq.(\ref{ham}) to the energy density is given as
\be
V_D=
- G_D \langle \Omega(\beta,\mu)|
(\bar\psi\gamma^5\epsilon^b\psi^C)(\bar\psi^C\gamma^5\epsilon\epsilon^b\psi)
| \Omega(\beta,\mu)\rangle
=-16G_DI_D^2
\label{vd}
\ee
with $I_D$ defined in Eq.(\ref{id})  
and the contribution from the determinant term is as given
in Eq.(\ref{vdet}).

To calculate the thermodynamic potential we shall have to specify the
chemical potentials relevant for the system.
Here we shall be interested in the form of quark matter that might be present
in compact stars older than few minutes so that 
chemical equilibration under weak
interaction is there. The relevant chemical potentials in this case 
are the baryon chemical potential $\mu_B(=3\mu_q)$, the chemical potential 
$\mu_E$ associated with electromagnetic charge,
and the two color electrostatic chemical potentials $\mu_3$ and $\mu_8$.
The chemical potential
is a matrix that is diagonal  in color and flavor space, and is given by
\be
\mu_{ij,ab}=(\mu\delta_{ij}+Q_{ij}\mu_E)\delta_{ab}
+(Q_{3ab}\mu_3+Q_{8ab}\mu_8)\delta_{ij}.
\label{muij}
\ee
Demanding the color superconducting
ground state to be invariant under  $SU(2)_c$ gauge group, we can
choose the chemical potential $\mu_3$ to be zero.

The total thermodynamic potential, including the contribution from the
electrons, is then given by
\be
\Omega=T+V_S+V_D-\langle \mu N\rangle-\frac{1}{\beta}s+\Omega_e
\label{Omega}
\ee
where, we have denoted
\be
\langle \mu N\rangle=\langle \psi^{ia\dagger}\mu_{ij,ab}\psi^{jb}\rangle
=2\sum_{i,a}\mu^{ia}I_v^{ia}.
\ee
In the above, $\mu^{ia} $ is the chemical potential for the quark of flavor $i$
and color $a$, which can be expressed in terms of the chemical potentials
$\mu$, $\mu_E$ and $\mu_8$ using Eq.(\ref{muij}). 
Further
\be
I_v^{ia}=\frac{1}{(2\pi)^3}\int d\zbf k(F^{ia}-F_1^{ia})
\label{iv}
\ee
is proportional to the number density of quarks of a given color
and flavor. The thermodynamic potential for electrons is given as
\be
\Omega_e=
-\frac{\mu_E^4}{12\pi^2}\left(1+2\pi^2 \frac{T^2}{\mu_E^2}\right).
\label{omge}
\ee
Here, the electron mass is assumed  to be zero which suffices for
the system we are considering.

Finally, for the entropy density for the quarks we have \cite{tfd}
\bearr
s & = & -\frac{2}{(2\pi)^3}\sum_{i,a}\int d \zbf k
\Big ( \sin^2\theta^{ia}_-\ln \sin^2\theta^{ia}_-
+\cos^2\theta^{ia}_-\ln \cos^2\theta^{ia}_- \nonumber \\ 
& + & \sin^2\theta^{ia}_+\ln \sin^2\theta^{ia}_+
+\cos^2\theta^{ia}_+\ln \cos^2\theta^{ia}_+\Big ).
\label{ent}
\eearr

Now the functional  minimisation of  the thermodynamic potential
 $\Omega$ with respect to  the chiral condensate function 
$h _i (\zbf k)$ leads to
\be
\cot \phi_i(\zbf k)
= \frac{ m_i+8G_sI_s^{(i)}+8K|\epsilon^{ijk}|Is^{(j)}I_s^{(k)}
+4KI_D^2\delta^{i3}}{|\zbf k|}
\equiv \frac{M_i}{|\zbf k|}.
\label{tan2h}
\ee
\noindent The last term in the numerator indicates the
explicit dependence of the strange quark--antiquark condensate function on the
light diquark condensate $I_D$.

Substituting this back in Eq.(\ref{is}) yields the mass gap equation as
\be
M_i=m_i+ \frac{ 8G_s}{(2\pi)^3}
\int \frac{M_i}{\epsilon_i^2} \sum_{a=1,3}(1-F^{ia}-F^{ia}_1)
d \zbf k +8K|\epsilon_{ijk}|I_s^{(j)}I_s^{(k)}+4KI_D^2\delta^{i3},
\label{mgap}
\ee
with, $\epsilon_i=\sqrt{\zbf k^2+M_i^2}$ being the energy of the
constituent quarks of i-th flavor. Note that for the two flavor
superconductivity as considered here, the strange quark mass is 
affected explicitly by the superconducting gap given by
the last term on the right hand side Eq.(\ref{mgap}). Of course, there is
implicit dependence on the gap in the second term through the functions
 $F$ and $F_1$ (given in Eq.s (\ref{fkb}) and (\ref{f1kb})).
Further, when chiral symmetry is restored for the light
quarks i.e., when the scalar condensate for the nonstrange quarks vanishes, 
still, the determinant term gives rise to a  density dependent
dynamical strange quark mass. Such a mass generation
is very different from the typical mechanism of quark mass generation
through quark--antiquark condensates \cite{steiner}.

Next, the minimisation of the thermodynamic potential  $\Omega$ 
with respect to the diquark  and di-antiquark condensate functions 
$f(\zbf k)$ and $f_{1}(\zbf k)$ yields 

\be
\tan 2f(\zbf k)=\frac{4(2G_D+KI_s^{(3)})}{\bar \epsilon-\bar \mu}\equiv
\frac{\Delta}{\bar \epsilon-\bar \mu}
\cos(\frac{\phi_1-\phi_2}{2})
\label{tan2f}
\ee
and
\be
\tan 2f_1(\zbf k)=\frac{4(2G_D+KI_s^{(3)}I_D)}{\bar \epsilon+\bar \mu}\equiv
\frac{\Delta}{\bar \epsilon+\bar \mu}
\cos(\frac{\phi_1-\phi_2}{2})
\label{tan2f1}
\ee
where, we have defined the superconducting gap 
$\Delta= 4 (2G_D-KI_s^{(3)})I_D$, with
$I_D$ as defined in Eq.(\ref{id}). Further,
in the above $\bar\epsilon=(\epsilon_1+\epsilon_2)/2$, 
$\bar\mu= (\mu_{ur}+\mu_{dg})/2 = (\mu_{ug}+\mu_{dr})/2 =\mu+\mu_E/6+
\mu_8/\sqrt{3}$.
It is thus seen that the diquark condensate functions depend upon
the {\em average} energy and the {\em average} chemical potential
of the quarks that condense. We also note here that the 
diquark condensate functions depends upon the masses of the two quarks which
condense through the function $\cos \big ((\phi_1-\phi_2)/2\big )$.
The function $\cos\phi_i=M_i/\epsilon_i$, 
can be different for u,d quarks,
when the charge neutrality condition is imposed. 
Such a normalisation factor is always there when
the condensing fermions have different masses as has been noted in Ref.
\cite{aichlin} in the context of CFL phase. 

Substituting  the solutions for the condensate functions given in Eq.s
(\ref{tan2h}), (\ref{tan2f}) and (\ref{tan2f1}) in the expression for
$I_D$ in Eq.(\ref{id}), we have the superconducting gap equation given by
\be
\Delta=\frac{4(2G_D+KI_s^{3})}{(2\pi)^3}\int{d\zbf k}
\Delta\cos^2\Big(\frac{\phi_1-\phi_2}{2}\Big)
\left[\frac{1}{\bar\omega_-}\cos 2\theta_-
+\frac{1}{\bar\omega_+}\cos 2\theta_+\right].
\label{del}
\ee
In the above, 
 $\bar\omega_\pm
=\sqrt{\Delta^2\cos^2((\phi_1-\phi_2)/2) +\bar\xi_{\pm}^2}$,
${\bar \xi }_{\pm }=\bar \epsilon \pm {\bar \mu}$ ,and, $\cos 2\theta_\pm=
1-\sin^2 \theta_\pm^u-\sin^2\theta_\pm^d$.

Finally, the minimisation of the thermodynamic potential with respect to the
thermal functions $\theta_{\pm}(\zbf k)$ gives
\be
\sin^2\theta_\pm^{ia}=\frac{1}{\exp(\beta\omega_\pm^{ia})+1}.
\label{them}
\ee

 \noindent Various $\omega^{ia}$'s $(i,a\equiv {\rm {flavor,color}})$
are explicitly  given as
\begin{mathletters}
\be
\omega_\pm^{11} =
\omega_\pm^{12}=\bar\omega_\pm +\delta_\epsilon\pm \delta_\mu\equiv \omega_\pm^u
\label{omgpmu}
\ee

\be
\omega_\pm^{21} = \omega_\pm^{22} =
\bar\omega_\pm-\delta_\epsilon\pm\delta_\mu\equiv\omega_\pm^d
\label{omgpmd}
\ee
for the quarks participating in condensation, and,
\be
\omega_{\pm}^{ia}=\epsilon^i{\pm}\mu^{ia}
\label{disps}
\ee
\end{mathletters}
for the blue quarks which do not  participate in superconductivity.
\noindent Here
$\delta _ \epsilon$($=(\epsilon_1-\epsilon_2)/2$) is half the energy difference
of the two quarks which condense, and, $\delta_\mu=
(\mu_{11}-\mu_{22})/2=\mu_E/2$, is half the difference of the chemical potentials
of the two condensing quarks.
Further $\bar\omega_\pm$ have been already defined  after Eq.(\ref{del}).
 Note that when the charge neutrality conditions are not imposed,
 all the four quasi particles
taking part in diquark condensation  will have the
same energy $\bar\omega_-$. It is clear from the
dispersion relations given in Eq.(\ref{disps}) that  
it is possible to have zero modes, i.e., $\omega^{ia}=0$
depending upon the values of $\delta_\epsilon$
and $\delta_\mu$. So, although we shall have nonzero order
parameter $\Delta$, there will be fermionic zero modes or the 
gapless superconducting phase \cite {abrikosov, krischprl}. 

Now using these dispersion relations, the mass
gap equation Eq.(\ref{mgap}) and the superconducting gap
equation Eq.(\ref{del}), the thermodynamic potential 
(Eq.(\ref{Omega})) becomes
\be
\Omega=\Omega_{q}+\Omega_e.
\label{omgt}
\ee
In the above, $\Omega_{q}$ is the contribution from the  quarks
and is given as
\bearr
\Omega_{q}&=& \frac{8}{(2\pi)^3}
\int d\zbf k\left[\sqrt{\zbf k^2+m^2}-
\frac{1}{2}(\bar\omega_-+\bar\omega_+)\right]\nonumber\\
&-& \frac{4}{\beta(2\pi)^3}
\sum_{m=u,d}\int d\zbf k \left[\log(1+\exp (-\beta\omega_{-}^m)
+\log(1+\exp(-\beta\omega_{+}^m)\right]\nonumber\\
&+&\frac{4}{(2\pi)^3}
\int d\zbf k\left[\sqrt{\zbf k^2+m^2}-
\frac{1}{2}(\epsilon_1+\epsilon_2)\right]\nonumber\\
&-& \frac{2}{\beta(2\pi)^3}
\sum_{i=1,2}\int d\zbf k \left[\log(1+\exp({-\beta(\epsilon_i-\mu^{i3})})
+\log(1+\exp({-\beta(\epsilon_i+\mu^{i3})})\right]\nonumber\\
&+&
\frac{6}{(2\pi)^3}
\int d\zbf k\left[\sqrt{\zbf k^2+m_s^2}-
\sqrt{\zbf k^2+M_s^2}\right]\nonumber\\
&-& \frac{2}{\beta(2\pi)^3}
\sum_{a=1,3}\int d\zbf k \left[\log(1+\exp({-\beta(\epsilon_3-\mu^{3a})})
+\log(1+\exp({-\beta(\epsilon_3+\mu^{3a})})\right]\nonumber\\
&+& 16G_DI_D^2+8G_s\sum_{i=1,3}{I_s^{(i)}}^2+32KI_s^{(1)}I_s^{(2)}I_s^{(3)}
+16KI_s^{(3)}I_D^2,
\label{omgq}
\eearr
where, $\omega_{\pm}^{u,d}$ are given in equations (\ref{omgpmu}) and 
(\ref{omgpmd}).
The  contribution of the electron $\Omega_e$ to the total 
thermodynamic potential  is already given in Eq.(\ref{omge}).
The first two lines in Eq.(\ref{omgq}) correspond to the contributions
from the quarks taking part in the condensation while the 
third and fourth lines correspond to the contribution
from  the two light quarks with the blue color. The next two lines are 
the contributions essentially from the strange quarks. The last 
line corresponds to the terms which mix up the flavors 
and also the diquark and quark--antiquark condensates. In fact, the last two
terms in the expression of thermodynamic potential 
are the contributions arising due to the determinant interaction.


Thus the thermodynamic potential is a function of four parameters: the three
mass gaps and a superconducting gap. These are calculated through minimisation
of the thermodynamic potential,
subjected to the conditions of electrical and color charge neutrality. 
The electric and color charge neutrality constraints are given as
\be
Q_E=\frac{2}{3}\rho^1-\frac{1}{3}\rho^2-\frac{1}{3}{\rho^3}
-\rho_e=0,
\label{qe}
\ee
and,
\be
Q_8 =\frac{1}{\sqrt{3}}\sum_ i (\rho^{i1} + \rho ^{i2}
- 2 \rho ^{i3}) =0,
\label{q8}
\ee
respectively.
In the above $\rho^{ia}=\langle{\psi^{ia}}^\dagger\psi^{ia}\rangle
=2 I_v^{ia}$ ($i$, $a$ not summed) and $I_v^{ia}$ is as given in Eq.(\ref{iv}).
Further, $\rho^i=\sum_{a=1,3}\rho^{ia}$.
%
%
 \noindent The thermodynamic potentials (\eqrftw{omgt}{omgq}),
the charge neutrality conditions (\eqrftw{qe}{q8}), the mass gap equation 
(Eq.(\ref{mgap})) and the superconducting 
gap equation (Eq.(\ref{del})) constitute the basis of the numerical 
calculations that we shall discuss in the next section.
\section{Results and discussions}
\label{results}
For numerical calculations we have taken the values of
the parameters of the NJL model as follows. The coupling constants
$G_s$, $G_D$ have the dimensions of $[{\rm Mass}]^{-2}$ while 
the six fermion coupling
has a dimension $[{\rm Mass}]^{-5}$. To regularise the divergent integrals
we use a sharp cut-off, $\Lambda$ in 3-momentum space. Thus we have all
together six parameters, namely the current quark masses
for the nonstrange and strange quarks, $m_q$ and $m_s$,
the three couplings $G_s$, $G_D$, $K$ and the three-momentum cutoff $\Lambda$.
For simplicity, we shall also take $G_D$ to be 0.75$G_s$, as may be
expected from Fierzing a current current interaction. For the rest of the
parameters we choose $\Lambda=0.6023$ GeV, $G_s\Lambda^2=1.835$,
$K\Lambda^5=12.36$, $m_q=5.5$ MeV and $m_s=0.1407$ GeV  as has been 
used in Ref. \cite{rehberg}. After choosing $m_q=5.5$ MeV, the remaining four
parameters are fixed by fitting to the pion decay constant and
the masses of pion, kaon and $\eta'$. With this set of parameters the
mass of $\eta$ is underestimated by about six percent. 
With this parametrization, the constituent masses of the
light quarks turn out to be $M_1=0.368$ GeV 
for u-d quarks, while the same for strange quark comes out as $M_s=0.549$ GeV,
at zero temperature and zero density. 

It is, however, relevant here to
comment regarding the choice of the parameters.
There have been different sets of
parameters by other groups also \cite{hatkun,lkw,bubrep} for
the three flavor NJL model. Although the same
principle  as above is used e.g. in Ref \cite{hatkun}, the resulting
parameter sets are not identical -- in particular, the dimensionless
coupling $K\Lambda^5$ differs by as large as 30 percent as compared to
the value used here.  This discrepancy lies on different treatment
of the $\eta' $ meson.  Since NJL model does not confine, and because
of the large mass of the $\eta'$ meson ($m_{\eta'}$=958 MeV), it lies
above the threshold for $q\bar q $ decay with an unphysical
imaginary part for the corresponding polarization diagram. 
This is an unavoidable feature of NJL model and leaves uncertainty
which is reflected in the difference in the parameter sets by 
different groups. Within this limitation regarding the parameters 
of the model, however, we proceed with the above parameter set 
which has  already been used in the study of the phase diagram of 
dense matter in Ref. \cite{ruester} as well as in the context of 
equation of state for neutron star matter in Ref. \cite{leupold}.

Let us begin with the discussions of results 
when the charge neutrality conditions are not imposed. At zero temperature,
the behaviour of the gap parameters as functions of quark chemical potential
are displayed in Fig.1-a.
We may point out that these solutions for the gaps correspond 
to the solutions for which the thermodynamic potential is minimised. In fact, 
in general, for certain values of the chemical potential, there can be
several solutions of the gap equations particularly near the
critical chemical potential. We have
chosen the ones which minimise the thermodynamic potential. 
\begin{figure}
\vspace{-0.4cm}
\begin{center}
\begin{tabular}{c c }
\includegraphics[width=8cm,height=8cm]{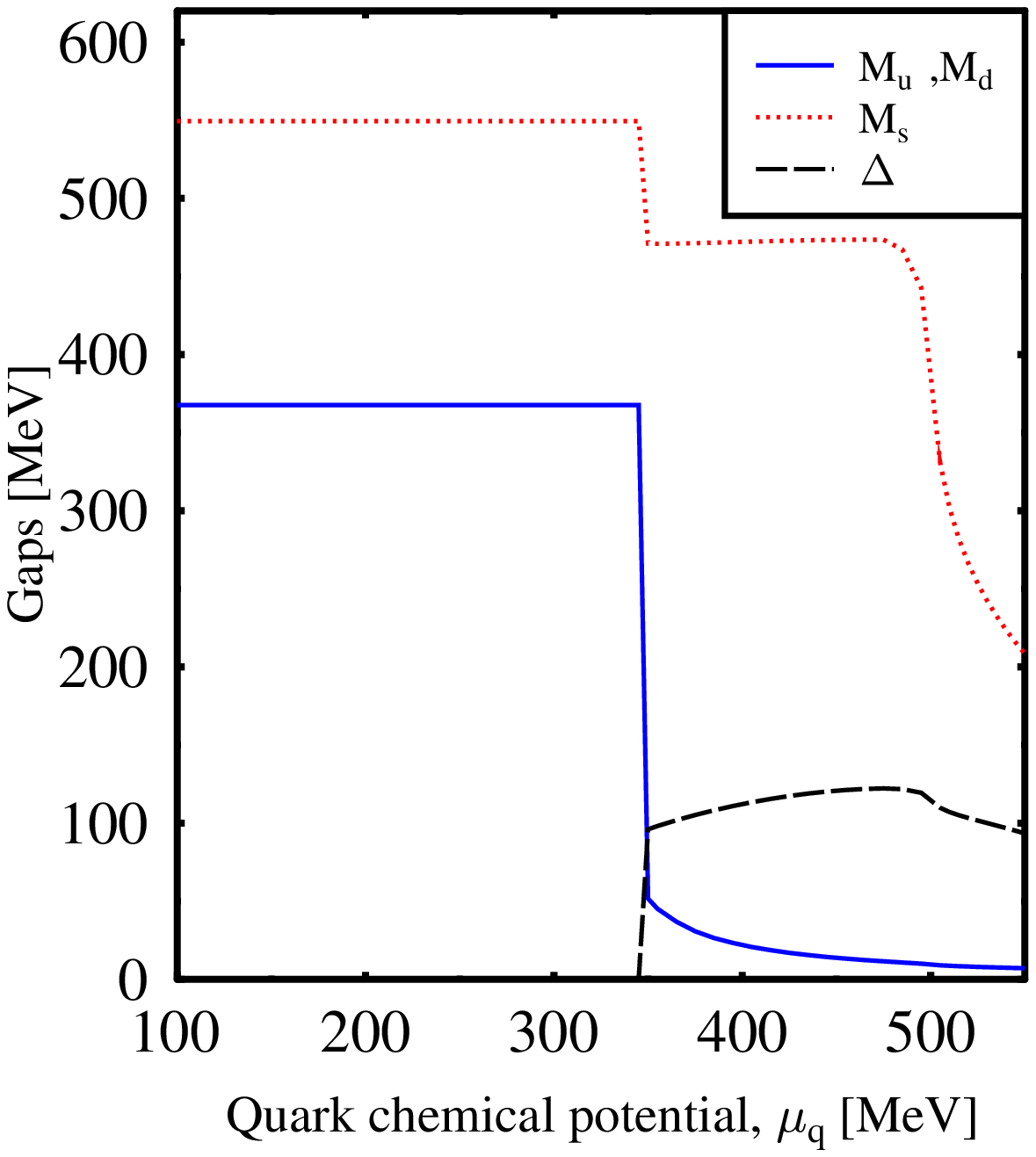}&
\includegraphics[width=8cm,height=8cm]{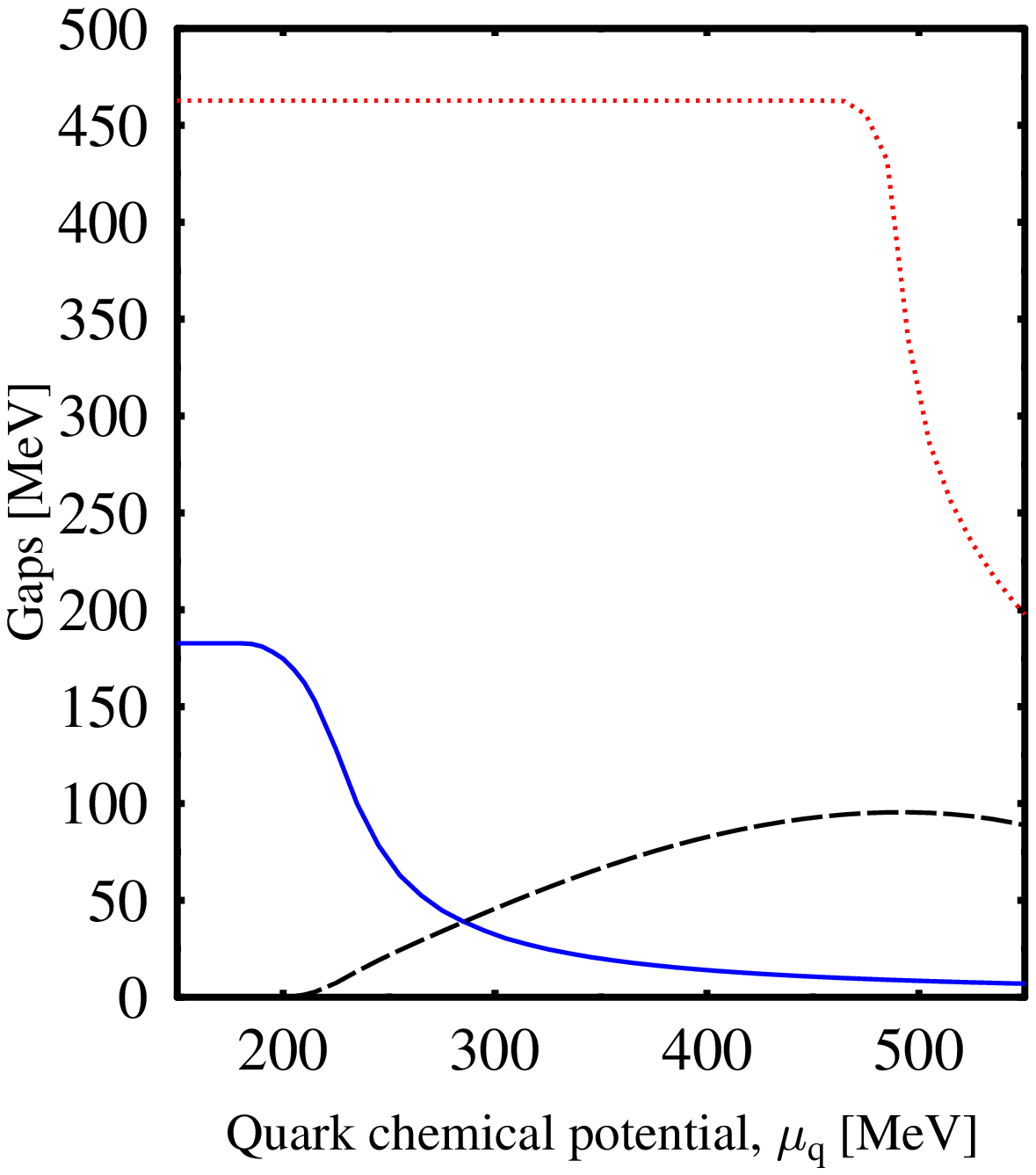}\\
Fig. 1-a & Fig.1-b
\end{tabular}
\end{center}
\caption{\it Gap parameters when charge neutrality conditions are not imposed.
Fig.1-a shows the gaps at zero temperature as a function of quark chemical
potential. Fig. 1-b  shows the same when the determinant interaction is not 
taken into account.
Both the figures correspond to nonzero value for the current quark
masses $m_u$=5.5 MeV=$m_d$.
Solid curve refers to masses of u and d quarks, the dotted
curve refers to  mass of strange quarks and the dashed curve corresponds to 
the superconducting gap.}
\label{fig1}
\end{figure}

At low chemical potentials $\mu_q<\mu_1\sim$ 350 MeV, the diquark gap vanishes
and the masses of the quarks stay at their vacuum values. The entire region 
below $\mu_q=\mu_1$ corresponds to vacuum solution and has zero baryon number.
At $\mu_q=\mu_1$ a first order phase transition takes place and the
system is a two flavor color superconductor. The diquark gap jumps from 
zero to about 95 MeV at this point. At the same point, the masses 
of u and d quarks drop from their vacuum values of 370 MeV to 50 MeV.
The baryon number density also jumps from zero to 0.42 fm$^{-3}$ 
at this chemical potential.
Because of the flavor mixing six Fermi interaction,
this cross over transition for the light quarks is reflected also
as dropping of the strange quark mass at this chemical potential from its 
vacuum value of 549 MeV to about 470 MeV.
Although this transition is not a first order transition due to the
nonzero current quark masses, there exist metastable phases.
The masses of the light quarks in these metastable phases
are the nontrivial solutions of the mass gap equations, but, have higher 
thermodynamic potential as compared to the solutions corresponding to
stable phases which are shown in Fig.1-a.
With the increase in $\mu_q$, the superconducting gap increases 
until it reaches a maximum at $\mu_q\sim$ 475 MeV. 
Beyond this point, as the quark  chemical potential is increased,
the mass of strange quark drops again. Due to the determinant interaction
this leads to a drop in superconducting gap as may be obvious from
Eq.(\ref{del}).
For comparison we have plotted  in Fig. 1-b 
the masses and the superconducting gap for the case
where the determinant coupling is zero, while all other couplings
remain unchanged. In this case, it may be noted
 that  the (nearly) vanishing of light quark masses
does not have any effect on the strange quark mass. The  strange quark
mass starts to drop only when quark chemical potential is larger
than the vacuum  value of the mass of the strange quark 
as only then the finite density contributions become nonzero 
in the mass gap equation for the strange quark.
In contrast, with nonzero coupling $K$,
because of the flavor mixing, the dropping of strange quark mass
starts for quark chemical potentials smaller than 
the vacuum mass of the strange quark as shown in Fig. 1-a. Since the charge
neutrality condition depends very sensitively on the strange quark mass,
determinant coupling therefore can have important consequences in deciding 
the phase structure.

We would also like to note that, to evaluate the thermodynamic
potential, we have to solve  self consistently the three mass gap equations
 (Eq.(\ref{mgap})) and the superconducting gap equation (Eq.(\ref{del})). 
These equations are all coupled.
However, some simplification occurs for higher densities if we take the
current quark masses to be zero. This leads to solving only two coupled
gap equations - the strange quark mass gap equation and the superconducting 
gap equation for densities when chiral symmetry is restored for the light
 u,d quarks. Although the nature of chiral symmetry transition changes
from a cross over to a first order transition, the values of the 
constituent masses
of the quarks do not change very much. Henceforth we shall limit 
our discussions to this case when current quark masses for the light quarks 
are taken to be zero. In Fig.2
we show the results of such a calculation for zero temperature when
the charge neutrality conditions are not imposed.
The general behaviour of the graphs
are similar to those in Fig. 1. The important difference is that, 
the chiral transition is a sharp first order transition at zero temperature. 
The constituent quark masses at
zero densities for up (down) and strange quarks are 354 MeV and 546 MeV
as compared to 368 MeV and 549 MeV respectively, when current quark masses
of u,d quarks are taken to be nonzero.
The critical chemical potential
turns out to be 335 MeV for the case of nonzero K. We also note
that as the quark chemical potential increases beyond 480 MeV,
the superconducting gap starts decreasing as the strange quark--antiquark
condensate decreases so that the effective diquark
coupling also decreases, as may be clear from Eq.(\ref{del}).

There have been calculations where the effect of the determinant interaction
is retained for mass gap calculations, but the effect of such an interaction 
is not included in the superconducting gap equation \cite{reddy,bubmix,ruester}.
We plot in Fig. 3 the superconducting gap including the effect of determinant
interaction for the u-d superconductivity gap for the case when the charge
neutrality conditions are not imposed.
As may be seen, including
the effect of strange quark--antiquark condensate through the determinant
interaction  leads to an enhancement of the superconducting gap and this
can be as large as a twenty five percent for chemical potentials of about 400 MeV.
This is because the `effective' coupling for superconductivity increases from
$G_D$ to $G_D-K \langle\bar s s\rangle/4$. At higher densities
however, this effect diminishes as the maginitude of the quark--antiquark 
condensate itself decreases. Such effects can play an
important role in deciding which phase can occur at what density as we
come down in the density. In particular, such kind of enhancement will be
there for u-d superconductivity, but will be absent for u-s or d-s
superconducting gaps since the light quark--antiquark condensate vanishes
much earlier.

\begin{figure}
\vspace{-0.4cm}
\begin{center}
\begin{tabular}{c c }
\includegraphics[width=8cm,height=8cm]{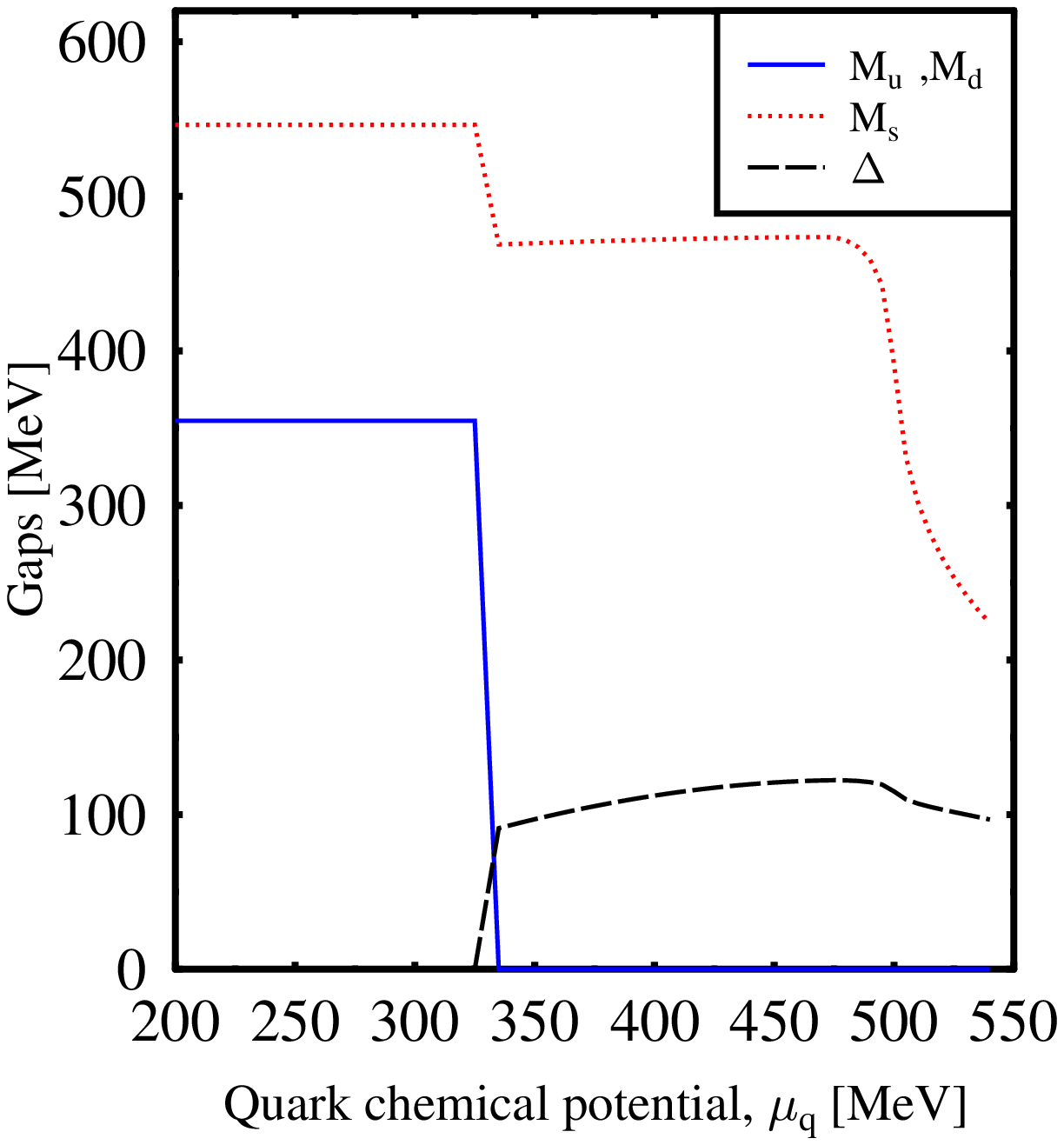}&
\includegraphics[width=8cm,height=8cm]{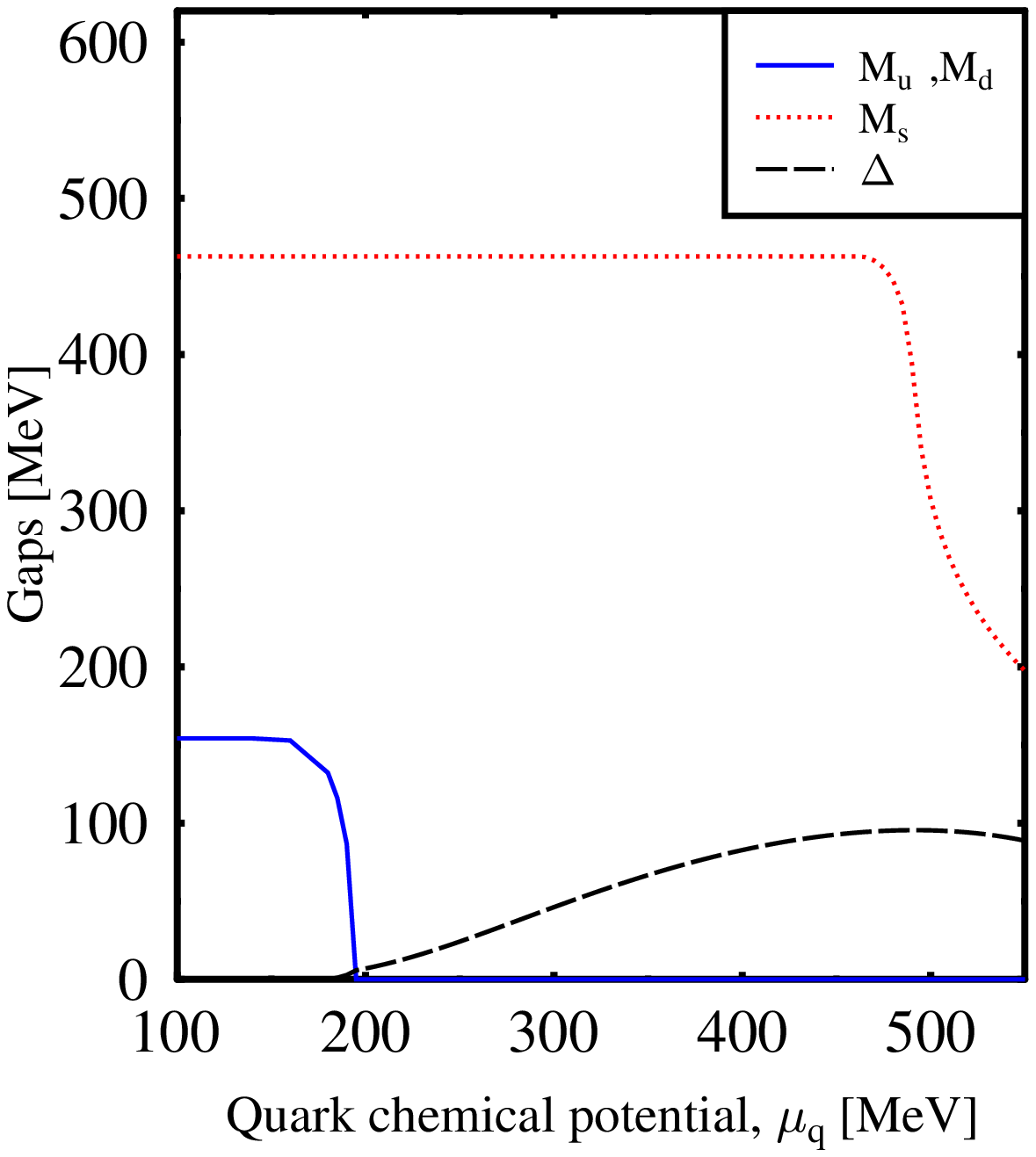}\\
Fig. 2-a & Fig.2-b
\end{tabular}
\end{center}
\caption{\it Same as in Fig. 1, but with zero current quark masses}
\label{fig2}
\end{figure}

\begin{figure}[htbp]
\begin{center}
\includegraphics[width=8cm,height=8cm]{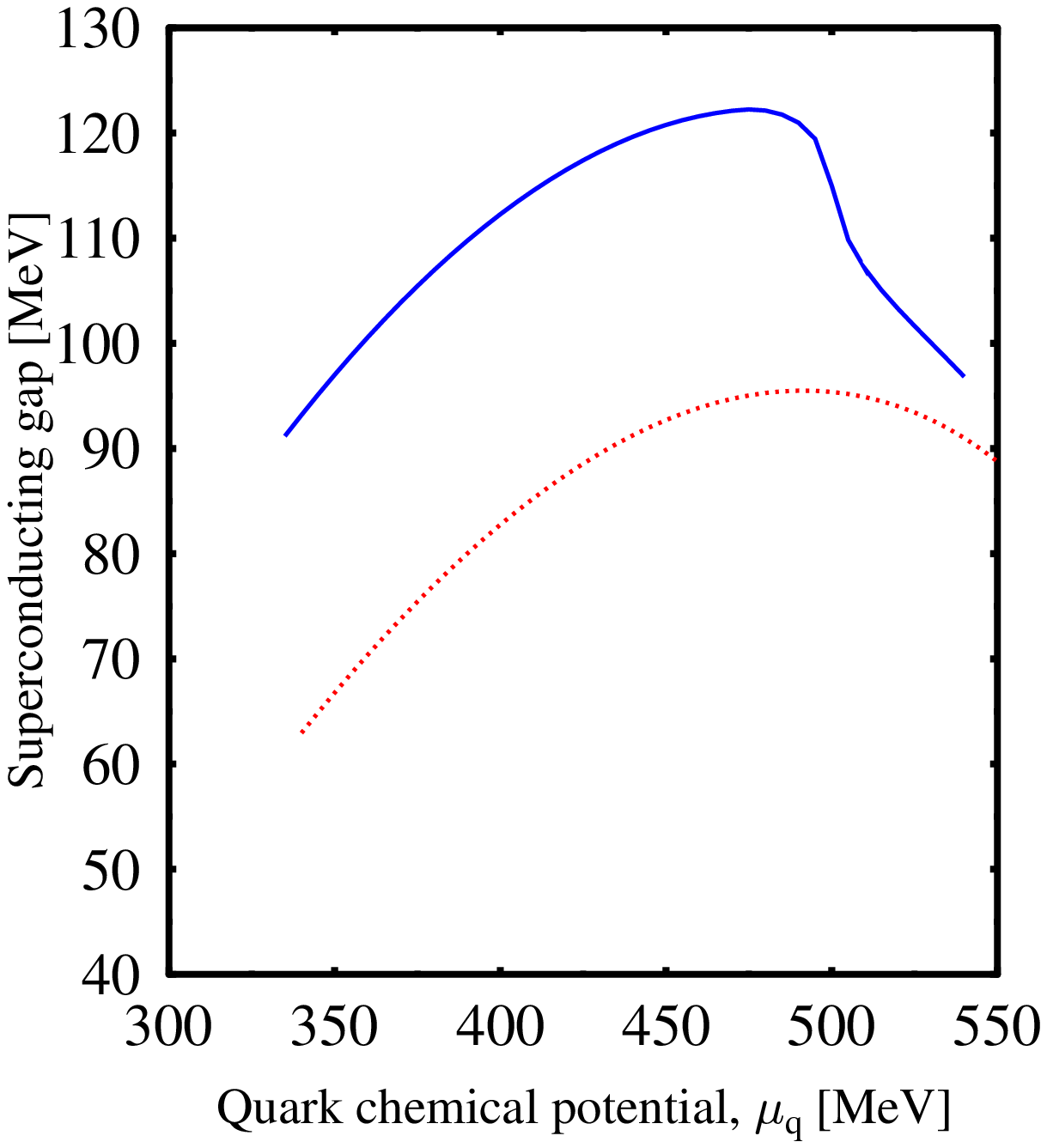}
\end{center}
\caption{\em{ Superconducting gap as a function of
quark chemical potential. Solid line corresponds to
including  the effect of strange quark--antiquark
condensate.  The dashed line corresponds to the case
when this effect is not included.}}
\label{gapk}
\end{figure}

\begin{figure}[htbp]
\begin{center}
\includegraphics[width=8cm,height=8cm]{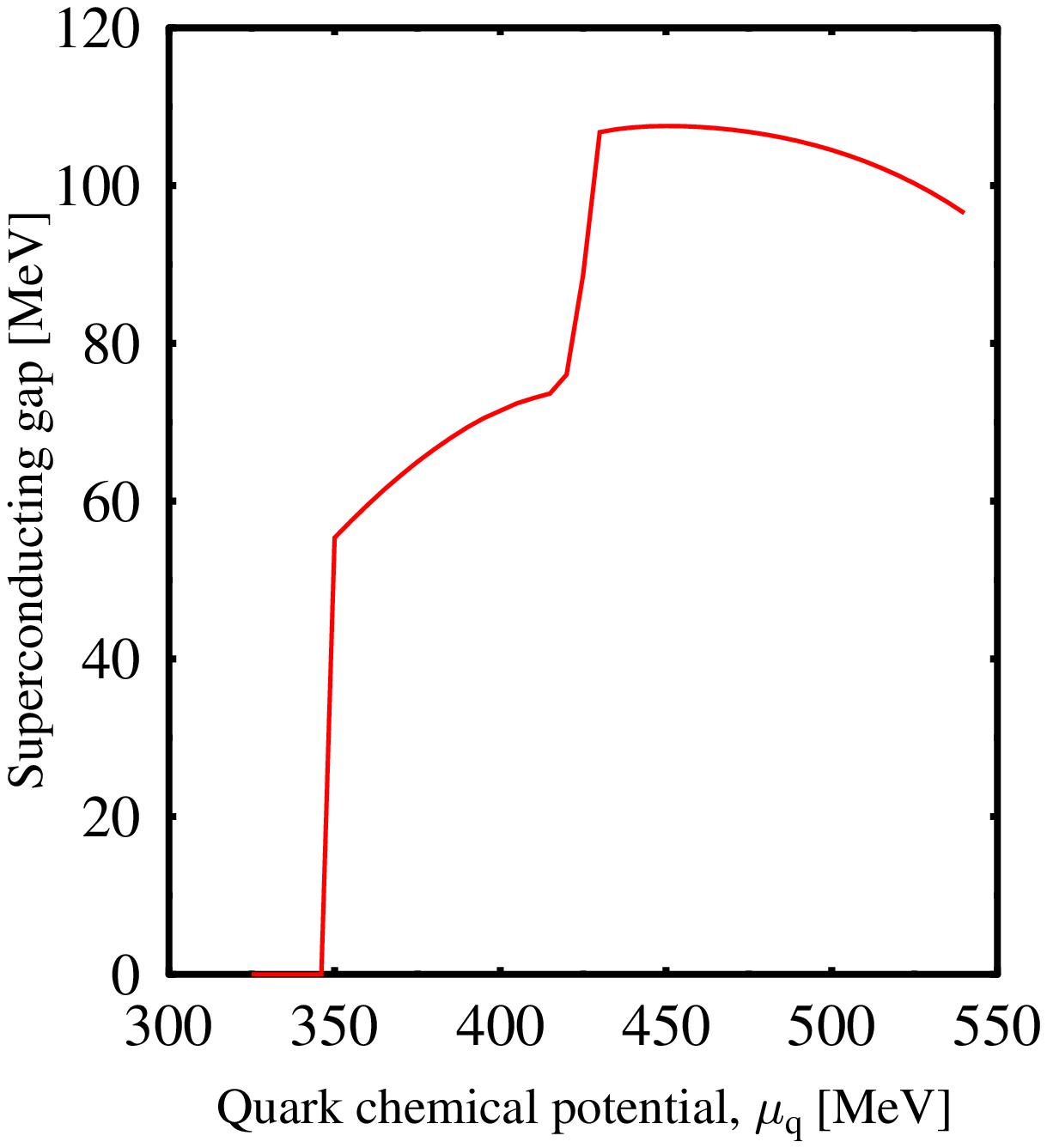}
\end{center}
\caption{\em{ Superconducting gap as a function of
quark chemical potential for the charge neural quark matter 
}}
\label{gapq0}
\end{figure}

We next extend our discussion to the case when the charge neutrality 
conditions ($Q_E$=0=$Q_8$) are imposed. 
We compute the thermodynamic potential numerically as follows.
For  given values of the quark chemical
potential, $\mu_q$, and the electric and color charge potentials,
$\mu_E$ and $\mu_8$, the  coupled mass gap equations and the superconducting 
gap equations are solved self consistently. The values of $\mu_E$ and $\mu_8$ 
are varied so that the charge neutrality 
conditions (Eq.(\ref{qe}) and Eq.(\ref{q8})) are satisfied. The resulting
solutions are then used in Eq.(\ref{omgq}) to compute the thermodynamic
potential. In doing so, we also check for existence of multiple solutions
of the gap equation and if they exist, we choose the solution which has
the least value for the thermodynamic potential.
In Fig.\ref{gapq0},  
we show the dependence of the superconducting gap on quark chemical potential
when charge neutrality condition is imposed. The superconducting gap
starts becoming nonzero for $\mu_q$ greater than 350 MeV. At this point the
chiral symmetry for the light quarks is restored. This has also its effect 
on the strange quark mass which drops from its vacuum value of about
546 MeV to 470 MeV.
The superconducting gap increases smoothly with $\mu_q$ until 
$\mu_q$ attains a value of around 
430 MeV. At this point the gap jumps from 80 MeV to 106 MeV and 
then increases slowly upto a maximum value of about 107.5 MeV at 
$\mu_q=450$ MeV. Beyond this value of $\mu_q$, the magnitude of the
strange quark--antiquark condensate decreases leading to a drop of 
superconducting gap through the determinant interaction.

As the quark chemical potential increases, strange quarks help in maintaining 
the charge neutrality conditions. We might observe here that
the strange quark mass starts decreasing already at $\mu_q=425$ MeV,
when charge neutrality conditions are imposed. This may be compared
with $\mu_q\simeq 480$ MeV, when charge neutrality conditions are not imposed
(see Fig. 2a). As we have already mentioned, there could be multiple 
solutions of the gap equation and whichever has the least free energy 
is the stable solution.
It could so happen that a solution of the mass gap equation which is metastable
when charge neutrality conditions are not imposed, can be the only solution
when charge neutrality conditions are imposed.

\begin{figure}[htbp]
\begin{center}
\includegraphics[width=8cm,height=8cm]{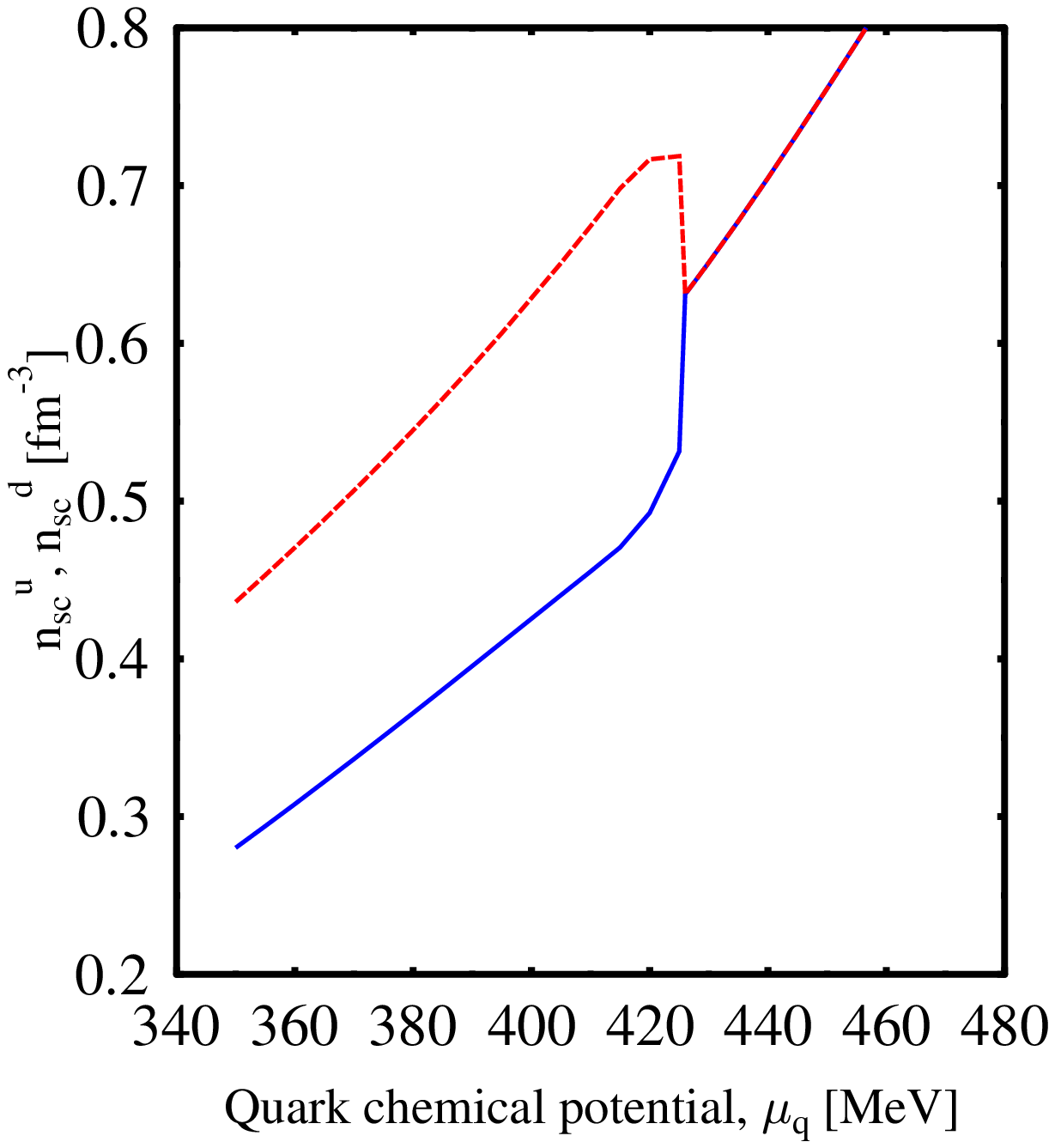}
\end{center}
\caption{\em Number densities  of u quarks (solid) and d quarks (dot-dashed) 
participating in superconducting phase.}
\label{rhoud}
\end{figure}

The pairing between the quarks with charge neutralty conditions correspond 
to stressed pairing i.e. the pairing of the quarks of different species 
which differ in their Fermi momenta \cite{andreaskris}. This gives rise 
to possibility of a gapless superconducting phase,
the QCD analogue of Sarma phase \cite{sarma}. In the present case, 
between the quark chemical potentials
$\mu_q\simeq$ 350 MeV and $\mu_q\simeq$ 425 MeV, the system is in
the gapless phase.
The number densities of $u$ and $d$ quarks participating in the condensation
are plotted in Fig. \ref{rhoud}.
As the quark chemical potential is increased beyond 425 MeV 
the solution for the gap jumps to a higher value of
about 106 MeV almost similar to the case when charge neutrality
conditions are not imposed. This corresponds to the usual BCS
solution. In this case, 
the number densities  of u quarks and d quarks participating in the 
condensation are the same. The charge neutrality conditions however 
are maintained 
by the blue colored u,d quarks, the electron as well as the strange quarks.
One essential effect of including strange quarks is that the electron density
starts decreasing for chemical potentials greater than the strange quark
mass and the strange quarks carrying the negative charge maintain
electric charge neutrality condition. This has the effect that
the lowest excitation energy $\omega_2=\bar\omega+\delta_\mu$
in the BCS pairing case becomes large due to both the large
value of the gap as well as due to the small magnitude of the
electron chemical potential.

It may be worthwhile to mention here that the gapless modes in
superconductivity were known theoretically in the context of 
superconducting matter with finite momentum \cite{abrikosov} 
as well as in condensed matter systems with magnetic impurities 
\cite{sarma}. Recently it has been 
investigated for cold fermionic atoms \cite{wilczek,rapid}. 
Gapless modes in the
context of quark matter has been first proposed in Ref. \cite{krischprl} for
color flavor locked matter. However, this corresponded to a metastable phase.
Gapless modes in neutral quark matter were first emphasized
for the two flavor color superconductivity in Ref. \cite{igor}
and for the 2SC+s quark matter in Ref. \cite{hmam}.
Stable gapless modes for color flavor
locked matter were first proposed in Ref.\cite{kriscfl} and have 
been confirmed in Ref.  \cite{ruester} in a more general structure
for the gap. The temperature dependence of the CFL gapless modes
has been studied in Ref.\cite{krisaug}.  We might mention here that the effects
 of nonzero strange quark mass
and the charge neutrality condition on superconducting gaps have also been
studied within a Ginzberg Landau approach in Ref.\cite{ida}.

\section{Summary}
\label{summary}
We have analysed here the effect of flavor mixing t'Hooft six fermion interaction
both for chiral symmetry breaking as well as two flavor superconductivity
in NJL model. The mehod is a variational one with an explicit construct for
the trial state including quark--antiquark as well as diquark condensates. 
The determinant interaction affects explicitly both the 
u-d superconducting gap as well as the mass gap for the strange quarks.
For the strange quarks, a density dependent mass arises even when the 
quark--antiquark condensates vanish. Such type of generation of 
 dynamical mass is  entirely distinct
from the typical mechanism of spontaneous chiral symmetry breaking through
quark--antiquark condensates \cite{steiner}.
Further, the u-d superconducting gap gets enhanced as inclusion of such a
term effectively increases the diquark coupling by an amount proportional
to the strange quark--antiquark condensate.

We have focussed our attention here to the  two flavor superconducting phase,
with  unpaired strange quarks. 
The variational method adopted can be directly generalised 
to include color flavor locked 
phase and one can then make a free energy comparison
regarding the possibility of which phase would be thermodynamically favorable
at what density.
In case of CFL phase, such an enhancement of superconducting u-s (d-s)
condensates will not be there as the corresponding 
$\langle \bar d d\rangle$($\langle \bar u u\rangle$) will vanish much earlier
than $\langle \bar s s\rangle$ condensates, when the density is increased. 
Since strange quark mass is a sensitive parameter in maintaining 
the charge neutrality condition, this will be important while comparing
free energies of different phases of charge neutral quark matter.

%
\acknowledgments
One of the authors (AM) would like to thank Frankfurt Institute for 
Advanced Studies (FIAS) for warm hospitality,
where the present work was initiated and Alexander von Humboldt Foundation, 
Germany for financial support.

\def \ltg{R.P. Feynman, Nucl. Phys. B 188, 479 (1981); 
K.G. Wilson, Phys. Rev. \zbf  D10, 2445 (1974); J.B. Kogut,
Rev. Mod. Phys. \zbf  51, 659 (1979); ibid  \zbf 55, 775 (1983);
M. Creutz, Phys. Rev. Lett. 45, 313 (1980); ibid Phys. Rev. D21, 2308
(1980); T. Celik, J. Engels and H. Satz, Phys. Lett. B129, 323 (1983)}

\def\berges {J. Berges, K. Rajagopal, {\NPB{538}{215}{1999}}.}
\def \svz {M.A. Shifman, A.I. Vainshtein and V.I. Zakharov,
Nucl. Phys. B147, 385, 448 and 519 (1979);
R.A. Bertlmann, Acta Physica Austriaca 53, 305 (1981)}

\def \spmbst {S.P. Misra, Phys. Rev. D35, 2607 (1987)}

\def \hmgrnv { H. Mishra, S.P. Misra and A. Mishra,
Int. J. Mod. Phys. A3, 2331 (1988)}

\def \snss {A. Mishra, H. Mishra, S.P. Misra
and S.N. Nayak, Phys. Lett 251B, 541 (1990)}

\def \amqcd { A. Mishra, H. Mishra, S.P. Misra and S.N. Nayak,
Pramana (J. of Phys.) 37, 59 (1991). }
\def\qcdtb{A. Mishra, H. Mishra, S.P. Misra 
and S.N. Nayak, Z.  Phys. C 57, 233 (1993); A. Mishra, H. Mishra
and S.P. Misra, Z. Phys. C 58, 405 (1993)}

\def \spmtlk {S.P. Misra, Talk on {\it `Phase transitions in quantum field
theory'} in the Symposium on Statistical Mechanics and Quantum field theory, 
Calcutta, January, 1992, hep-ph/9212287}

\def \hmnj {H. Mishra and S.P. Misra, 
{\PRD{48}{5376}{1993}.}}

\def \hmqcd {A. Mishra, H. Mishra, V. Sheel, S.P. Misra and P.K. Panda,
hep-ph/9404255 (1994)}

\def \amcrl {A. Mishra, H. Mishra and S.P. Misra, Z. Phys. C 57, 241 (1993)}

\def \higgs { S.P. Misra, in {\it Phenomenology in Standard Model and Beyond}, 
Proceedings of the Workshop on High Energy Physics Phenomenology, Bombay,
edited by D.P. Roy and P. Roy (World Scientific, Singapore, 1989), p.346;
A. Mishra, H. Mishra, S.P. Misra and S.N. Nayak, Phys. Rev. D44, 110 (1991)}

\def \nmtr {A. Mishra, 
H. Mishra and S.P. Misra, Int. J. Mod. Phys. A5, 3391 (1990); H. Mishra,
 S.P. Misra, P.K. Panda and B.K. Parida, Int. J. Mod. Phys. E 1, 405, (1992);
 {\it ibid}, E 2, 547 (1993); A. Mishra, P.K. Panda, S. Schrum, J. Reinhardt
and W. Greiner, to appear in Phys. Rev. C}

\def \dtrn {P.K. Panda, R. Sahu and S.P. Misra, 
Phys. Rev C45, 2079 (1992)}

\def \qcd {G. K. Savvidy, Phys. Lett. 71B, 133 (1977);
S. G. Matinyan and G. K. Savvidy, Nucl. Phys. B134, 539 (1978); N. K. Nielsen
and P. Olesen, Nucl.  Phys. B144, 376 (1978); T. H. Hansson, K. Johnson,
C. Peterson Phys. Rev. D26, 2069 (1982)}

\def \cornwal {J.M. Cornwall, Phys. Rev. D26, 1453 (1982)}
\def\aichlin {F. Gastineau, R. Nebauer and J. Aichelin,
{\PRC{65}{045204}{2002}}.}

\def \mndglv {J. E. Mandula and M. Ogilvie, Phys. Lett. 185B, 127 (1987)}

\def \schwinger {J. Schwinger, Phys. Rev. 125, 1043 (1962); ibid,
127, 324 (1962)}

\def \schutte {D. Schutte, Phys. Rev. D31, 810 (1985)}

\def \amspm {A. Mishra and S.P. Misra, 
{\ZPC{58}{325}{1993}}.}

\def \gft{ For gauge fields in general, see e.g. E.S. Abers and 
B.W. Lee, Phys. Rep. 9C, 1 (1973)}

\def \gribov {V.N. Gribov, Nucl. Phys. B139, 1 (1978)}

\def \spm78 {S.P. Misra, Phys. Rev. D18, 1661 (1978); {\it ibid}
D18, 1673 (1978)} 

\def \lopr {A. Le Youanc, L.  Oliver, S. Ono, O. Pene and J.C. Raynal, 
Phys. Rev. Lett. 54, 506 (1985)}

\def \spphi {S.P. Misra and S. Panda, Pramana (J. Phys.) 27, 523 (1986);
S.P. Misra, {\it Proceedings of the Second Asia-Pacific Physics Conference},
edited by S. Chandrasekhar (World Scientific, 1987) p. 369}

\def\spmdif {S.P. Misra and L. Maharana, Phys. Rev. D18, 4103 (1978); 
    S.P. Misra, A.R. Panda and B.K. Parida, Phys. Rev. Lett. 45, 322 (1980);
    S.P. Misra, A.R. Panda and B.K. Parida, Phys. Rev. D22, 1574 (1980)}

\def \spmvdm {S.P. Misra and L. Maharana, Phys. Rev. D18, 4018 (1978);
     S.P. Misra, L. Maharana and A.R. Panda, Phys. Rev. D22, 2744 (1980);
     L. Maharana,  S.P. Misra and A.R. Panda, Phys. Rev. D26, 1175 (1982)}

\def\spmthr {K. Biswal and S.P. Misra, Phys. Rev. D26, 3020 (1982);
               S.P. Misra, Phys. Rev. D28, 1169 (1983)}

\def \spmstr { S.P. Misra, Phys. Rev. D21, 1231 (1980)} 

\def \spmjet {S.P. Misra, A.R. Panda and B.K. Parida, Phys. Rev Lett. 
45, 322 (1980); S.P. Misra and A.R. Panda, Phys. Rev. D21, 3094 (1980);
  S.P. Misra, A.R. Panda and B.K. Parida, Phys. Rev. D23, 742 (1981);
  {\it ibid} D25, 2925 (1982)}

\def \arpftm {L. Maharana, A. Nath and A.R. Panda, Mod. Phys. Lett. 7, 
2275 (1992)}

\def \van {R. Van Royen and V.F. Weisskopf, Nuov. Cim. 51A, 617 (1965)}

\def \rchpi {S.R. Amendolia {\it et al}, Nucl. Phys. B277, 168 (1986)}

\def \chrl{ Y. Nambu, {\PRL{4}{380}{1960}};
A. Amer, A. Le Yaouanc, L. Oliver, O. Pene and
J.C. Raynal,{\PRL{50}{87}{1983a}};{\em ibid}
{\PRD{28}{1530}{1983}};
M.G. Mitchard, A.C. Davis and A.J.
MAacfarlane, {\NPB{325}{470}{1989}};
B. Haeri and M.B. Haeri,{\PRD{43}{3732}{1991}};
V. Bernard,{\PRD{34}{1604}{1986}};
 S. Schramm and
W. Greiner, Int. J. Mod. Phys. \zbf E1, 73 (1992), 
J.R. Finger and J.E. Mandula, Nucl. Phys. \zbf B199, 168 (1982),
S.L. Adler and A.C. Davis, Nucl. Phys.\zbf  B244, 469 (1984),
S.P. Klevensky, Rev. Mod. Phys.\zbf  64, 649 (1992).}

\def \spmijp { S.P. Misra, Ind. J. Phys. 61B, 287 (1987)}

\def \feynman {R.P. Feynman and A.R. Hibbs, {\it Quantum mechanics and
path integrals}, McGraw Hill, New York (1965)}

\def \glstn{ J. Goldstone, Nuov. Cim. \zbf 19, 154 (1961);
J. Goldstone, A. Salam and S. Weinberg, Phys. Rev. \zbf  127,
965 (1962)}

\def \anderson {P.W. Anderson, Phys. Rev. \zbf {110}, 827 (1958)}

\def \nambu{ Y. Nambu, Phys. Rev. Lett. \zbf 4, 380 (1960)}

\def\donogh {J.F. Donoghue, E. Golowich and B.R. Holstein, {\it Dynamics
of the Standard Model}, Cambridge University Press (1992)}

\def\satz {T. Matsui and H. Satz, Phys. Lett. B178, 416 (1986)}

\def\cps {C. P. Singh, Phys. Rep. 236, 149 (1993)}

\def\prliop {A. Mishra, H. Mishra, S.P. Misra, P.K. Panda and Varun
Sheel, Int. J. of Mod. Phys. E 5, 93 (1996)}

\def\hmcor {V. Sheel, H. Mishra and J.C. Parikh, Phys. Lett. B382, 173
(1996); {\it biid}, to appear in Int. J. of Mod. Phys. E}
\def\cort { V. Sheel, H. Mishra and J.C. Parikh, Phys. ReV D59,034501 (1999);
{\it ibid}Prog. Theor. Phys. Suppl.,129,137, (1997).}

\def\surcor {E.V. Shuryak, Rev. Mod. Phys. 65, 1 (1993)} 

\def\stevenson {A.C. Mattingly and P.M. Stevenson, Phys. Rev. Lett. 69,
1320 (1992); Phys. Rev. D 49, 437 (1994)}

\def\mac {M. G. Mitchard, A. C. Davis and A. J. Macfarlane,
 Nucl. Phys. B 325, 470 (1989)} 
\def\tfd
 {H.~Umezawa, H.~Matsumoto and M.~Tachiki {\it Thermofield dynamics
and condensed states} (North Holland, Amsterdam, 1982) ;
P.A.~Henning, Phys.~Rep.253, 235 (1995).}
\def\amph4{Amruta Mishra and Hiranmaya Mishra,
{\JPG{23}{143}{1997}}.}

\def \neglecor{M.-C. Chu, J. M. Grandy, S. Huang and 
J. W. Negele, Phys. Rev. D48, 3340 (1993);
ibid, Phys. Rev. D49, 6039 (1994)}

\def\revdata {Particle Data Group, Phys. Rev. D 50, 1173 (1994)}

\def\sinp {S.P. Misra, Indian J. Phys., {\bf 70A}, 355 (1996)}
\def\hmparikh{H. Mishra and J.C. Parikh, {\NPA{679}{597}{2001}.}}
\def\krisch {M. Alford and K. Rajagopal, JHEP 0206,031,(2002)}
\def\reddy {A.W. Steiner, S. Reddy and M. Prakash,
{\PRD{66}{094007}{2002}.}}
\def\hmam {Amruta Mishra and Hiranmaya Mishra,
{\PRD{69}{014014}{2004}.}}
\def\hmampp {Amruta Mishra and Hiranmaya Mishra,
in preparation.}
\def\bryman {D.A. Bryman, P. Deppomier and C. Le Roy, Phys. Rep. 88,
151 (1982)}
\def\thooft {G. 't Hooft, Phys. Rev. D 14, 3432 (1976); D 18, 2199 (1978);
S. Klimt, M. Lutz, U. Vogl and W. Weise, Nucl. Phys. A 516, 429 (1990)}
\def\alkz { R. Alkofer, P. A. Amundsen and K. Langfeld, Z. Phys. C 42,
199(1989), A.C. Davis and A.M. Matheson, Nucl. Phys. B246, 203 (1984).}
\def\sarah {T.M. Schwartz, S.P. Klevansky, G. Papp,
{\PRC{60}{055205}{1999}}.}
\def\wil{M. Alford, K.Rajagopal, F. Wilczek, {\PLB{422}{247}{1998}};
{\it{ibid}}{\NPB{537}{443}{1999}}.}
\def\sursc{R.Rapp, T.Schaefer, E. Shuryak and M. Velkovsky,
{\PRL{81}{53}{1998}};{\it ibid}{\AP{280}{35}{2000}}.}
\def\pisarski{
D. Bailin and A. Love, {\PR{107}{325}{1984}},
D. Son, {\PRD{59}{094019}{1999}}; 
T. Schaefer and F. Wilczek, {\PRD{60}{114033}{1999}};
D. Rischke and R. Pisarski, {\PRD{61}{051501}{2000}}, 
D. K. Hong, V. A. Miransky, 
I. A. Shovkovy, L.C. Wiejewardhana, {\PRD{61}{056001}{2000}}.}
\def\leblac {M. Le Bellac, {\it Thermal Field Theory}(Cambridge, Cambridge University
Press, 1996).}
\def\bcs{A.L. Fetter and J.D. Walecka, {\it Quantum Theory of Many
particle Systems} (McGraw-Hill, New York, 1971).}
\def\alexander{Aleksander Kocic, Phys. Rev. D33, 1785,(1986).}
\def\bubmix{F. Neumann, M. Buballa and M. Oertel,
{\NPA{714}{481}{2003}.}}
\def\kunihiro{M. Kitazawa, T. Koide, T. Kunihiro, Y. Nemeto,
{\PTP{108}{929}{2002}.}}
\def\igor{Igor Shovkovy, Mei Huang, {\PLB{564}{205}{2003}}.}
\def\prasanth{P. Jaikumar and M. Prakash,{\PLB{516}{345}{2001}}.}
\def\igorr{Mei Huang, Igor Shovkovy, {\NPA{729}{835}{2003}}.}
\def\abrikosov{A.A. Abrikosov, L.P. Gorkov, Zh. Eskp. Teor.39, 1781,
1960}
\def\krischprl{M.G. Alford, J. Berges and K. Rajagopal,
 {\PRL{84}{598}{2000}.}}
\def\hatmampp{A. Mishra and H.Mishra, in preparation}
\def\blaschke{D. Blaschke, M.K. Volkov and V.L. Yudichev,
{\EPJA{17}{103}{2003}}.}
\def\mei{M. Huang, P. Zhuang, W. Chao,
{\PRD{65}{076012}{2002}}}
\def\bubnp{M. Buballa, M. Oertel,
{\NPA{703}{770}{2002}}.}
\def\sarma{G. Sarma, J. Phys. Chem. Solids 24,1029 (1963).}
\def\ebert {D. Ebert, H. Reinhardt and M.K. Volkov,
Prog. Part. Nucl. Phys.{\bf 33},1, 1994.}
\def\rehberg{ P. Rehberg, S.P. Klevansky and J. Huefner,
{\PRC{53}{410}{1996}.}}
\def\lutz{M. Lutz, S. Klimt, W. Weise,{\NPA{542}{521}{1992}.}}
\def\rapid{B. Deb, A.Mishra, H. Mishra and P. Panigrahi,
Phys. Rev. A {\bf 70},011604(R), 2004.}
\def\kriscfl{M. Alford, C. Kouvaris, K. Rajagopal, Phys. Rev. Lett.
{\bf 92} 222001 (2004), arXiv:hep-ph/0406137.}
\def\shovris{S.B. Ruester, I.A. Shovkovy and D.H. Rischke,
arXiv:hep-ph/0405170.}
\def\krisaug{K. Fukushima, C. Kouvaris and K. Rajagopal, arxiv:hep-ph/0408322}.
\def\wilczek{W.V. Liu and F. Wilczek,{\PRL{90}{047002}{2003}},E. Gubankova,
W.V. Liu and F. Wilczek, {\PRL{91}{032001}{2003}.}}
\def\review{For reviews see K. Rajagopal and F. Wilczek,
arXiv:hep-ph/0011333; D.K. Hong, Acta Phys. Polon. B32,1253 (2001);
M.G. Alford, Ann. Rev. Nucl. Part. Sci 51, 131 (2001); G. Nardulli,
Riv. Nuovo Cim. 25N3, 1 (2002); S. Reddy, Acta Phys Polon.B33, 4101(2002);
T. Schaefer arXiv:hep-ph/0304281; D.H. Rischke, Prog. Part. Nucl. Phys. 52,
197 (2004); H.C. Ren, arXiv:hep-ph/0404074; M. Huang, arXiv: hep-ph/0409167;
I. Shovkovy, arXiv:nucl-th/0410191.}
\def\kunihiroo{ M. Kitazawa, T. Koide, T. Kunihiro and Y. Nemoto,
{\PRD{65}{091504}{2002}}, D.N. Voskresensky, arXiv:nucl-th/0306077.}
\def\rupak{S.Reddy and G. Rupak, arXiv:nucl-th/0405054}
\def\ida{K. Iida and G. Baym,{\PRD{63}{074018}{2001}},
Erratum-ibid{\PRD{66}{059903}{2002}}; K. Iida, T. Matsuura, M. Tachhibana 
and T. Hatsuda, {\PRL{93}{132001}{2004}}; ibid,{arXiv:hep-ph/0411356}}
\def\chromo{Mei Huang and Igor Shovkovy,{\PRD{70}{051501}{2004}};
 {\em ibid}, {\PRD{70}{094030}{2004}}}
\def\steiner{A.W. Steiner, {\PRD{72}{054024}{2005}.}}
\def\andreaskris{K. Rjagopal and A. Schimitt{\PRD{73}{045003}{2006}.}}
\def\amhm5{A. Mishra and H. Mishra, {\PRD{71}{074023}{2005}.}}
\def\leupold{K. Schertler, S. Leupold and J. Schaffner-Bielich,
{\PRC{60}{025801}(1999).}}
\def\bubrep{Michael Buballa, Phys. Rep.{\bf 407},205, 2005.}
\def\hatkun{T. hatsuda and T. Kunihiro, Phys. Rep.{\bf 247},221, 1994.}
\def\lkw{ M. Lutz, S. Klimt and W. Weise, Nucl Phys. {\bf A542}, 521, 1992.}
\def\ruester{S.B. Ruester, V.Werth, M. Buballa, I. Shovkovy, D.H. Rischke,
arXiv:nucl-th/0602018; S.B. Ruester, I. Shovkovy, D.H. Rischke,
{\NPA{743}{127}{2004}.}}

\end{document}